\documentclass{webofc}

\newcommand{\sNN}{\sqrt{s_{NN}}}

\usepackage[varg]{txfonts}   
\usepackage{hyperref}
\usepackage{url}
\hypersetup{colorlinks=true,citecolor=blue,urlcolor=blue,linkcolor=blue}
\begin{document}
\title{The collectivity of transverse momentum fluctuations}
%
%

\author{\firstname{Tribhuban} \lastname{Parida}\inst{1}\fnsep\thanks{\email{tribhu.451@gmail.com}} \and
        \firstname{Rupam} \lastname{Samanta}\inst{2}\fnsep\thanks{\email{rupam.samanta@ifj.edu.pl}} \and
        \firstname{Jean-Yves} \lastname{Ollitrault}\inst{3}\fnsep\thanks{\email{jean-yves.ollitrault@ipht.fr}}
}

\institute{Department of Physical Sciences,
Indian Institute of Science Education and Research Berhampur, Transit Campus (Govt ITI), Berhampur-760010, Odisha, India 
\and         
Institute of Nuclear Physics, Polish Academy of Sciences,  31-342 Cracow, Poland
\and
Universit\'e Paris Saclay, CNRS, CEA, Institut de physique th\'eorique, 91191 Gif-sur-Yvette, France
          }

\abstract{We study the observable $v_0(p_T)$, which quantifies the relative change of $p_T$ spectra induced by event-by-event density fluctuations in the medium created in heavy-ion collisions. This quantity provides a direct measure of radial flow and serves as a probe of collectivity, complementing anisotropic flow coefficients. Using hydrodynamic model calculations, we predict the behavior of $v_0(p_T)$ and show that the scaled quantity $v_0(p_T)/v_0$ exhibits very little dependence on centrality and transport coefficients. We further find that the apparent influence of transport coefficients$-$particularly bulk viscosity$-$ on $v_0(p_T)$ largely originates from modifications of the event-averaged mean transverse momentum, $\langle p_T \rangle$. By expressing $v_0(p_T)/v_0$ as a function of $p_T/\langle p_T \rangle$, the genuine sensitivity of $v_0(p_T)$ to transport coefficients can be isolated. Moreover, since $v_0(p_T)$ is the $p_T$-differential measure of event-by-event $[p_T]$ fluctuations, it naturally explains the observed $p_T$-cut dependence of $\sigma_{p_T}$ measured by ATLAS collaboration.}

\maketitle
\section{Introduction}
\label{intro}
In non-central heavy-ion collisions, the almond-shaped overlap region of the colliding nuclei leads to an anisotropy in the pressure gradients. This initial spatial anisotropy is transformed into a momentum-space anisotropy of the produced particles through the collective expansion of the medium, a phenomenon often referred to as shape-flow transmutation. The resulting anisotropy is quantified by flow coefficients. Among them, the observation of large elliptic flow in mid-central collisions stands out as one of the most striking manifestations of collectivity in heavy-ion collisions. Conventionally, flow coefficients are extracted from long-range correlations in the azimuthal angle of the produced particles. However, there also exists a long-range correlation that does not involve azimuthal angles, but rather appears in the transverse momentum of the produced particles, which has been comparatively less recognized as a signature of collectivity.

Events with same amount of deposited entropy can still exhibit different average transverse sizes due to event-by-event fluctuations in the positions of nucleons. More compact events, with smaller transverse size, lead to higher density and temperature, which in turn drive faster expansion and produce larger mean transverse momentum per particle, $[p_T]$, than less compact events. This effect is termed size-flow transmutation \cite{Bozek:2017elk}. In the hydrodynamic picture, compact events generate larger pressure gradients in the radial direction, which in turn result in stronger radial flow. At freeze-out, the momentum distribution of produced particles is determined by the fluid velocity distribution on the freeze-out hypersurface. Thus, variations in radial velocity directly modify the $p_T$ spectrum. Event-by-event fluctuations in transverse size therefore induce fluctuations in radial velocity, ultimately leading to variations in the $p_T$ spectra. To quantify this effect, the observable, $v_0(p_T)$, was introduced by Schenke, Shen, and Teaney \cite{Schenke:2020uqq}, which is defined as
\begin{equation}
\label{defv0pt}
v_{0}(p_T) \equiv \frac{\langle n(p_T)[p_T]\rangle-\langle n(p_T)\rangle\langle p_T\rangle}{\langle n(p_T)\rangle \sigma_{p_T}},
\end{equation}
where $n(p_T)$ denotes the fraction of particles in a given $p_T$ bin, $[p_T]$ is the event-wise mean transverse momentum, and $\langle p_T\rangle$ and $\sigma_{p_T}$ are the expectation value and standard deviation of $[p_T]$, respectively. For experimental measurements, it is recommended to evaluate $n(p_T)$ and $[p_T]$ in two rapidity windows separated by a gap in order to suppress non-flow contributions \cite{Parida:2024ckk}. The $p_T$-integrated $v_0$ is directly related to $\sigma_{p_T}$ through $v_{0} \equiv \sigma_{p_T}/\langle p_T\rangle$. Thus, $v_0(p_T)$ provides a $p_T$-differential measure of $\sigma_{p_T}$.

\section{Hydrodynamic model prediction}
\label{hydro}

Our first objective is to demonstrate that $v_0(p_T)$ originates from temperature fluctuations. To this end, we performed two types of hydrodynamic simulations using the MUSIC code for Pb+Pb collisions at $\sNN = 5.02$ TeV. In the first case, we carried out standard event-by-event simulations at a fixed impact parameter ($b$) and calculated $v_0(p_T)/v_0$ using the definition introduced earlier. In the second approach, we constructed a smooth Initial Condition (IC) by averaging over multiple TRENTo events, followed by hydrodynamic evolution of this smooth profile. The simulation with smooth IC was then repeated but with the entropy normalization increased by 5\%, resulting in a higher initial temperature. From this pair of simulations, we extracted $v_0(p_T)/v_0 = \frac{\delta \ln n(p_T)}{\delta \ln [p_T]},$ which represents the ratio between the logarithmic change of $n(p_T)$ and that of $[p_T]$ \cite{Parida:2024ckk}.

\begin{figure}[h]
\centering
\includegraphics[width=10cm,clip]{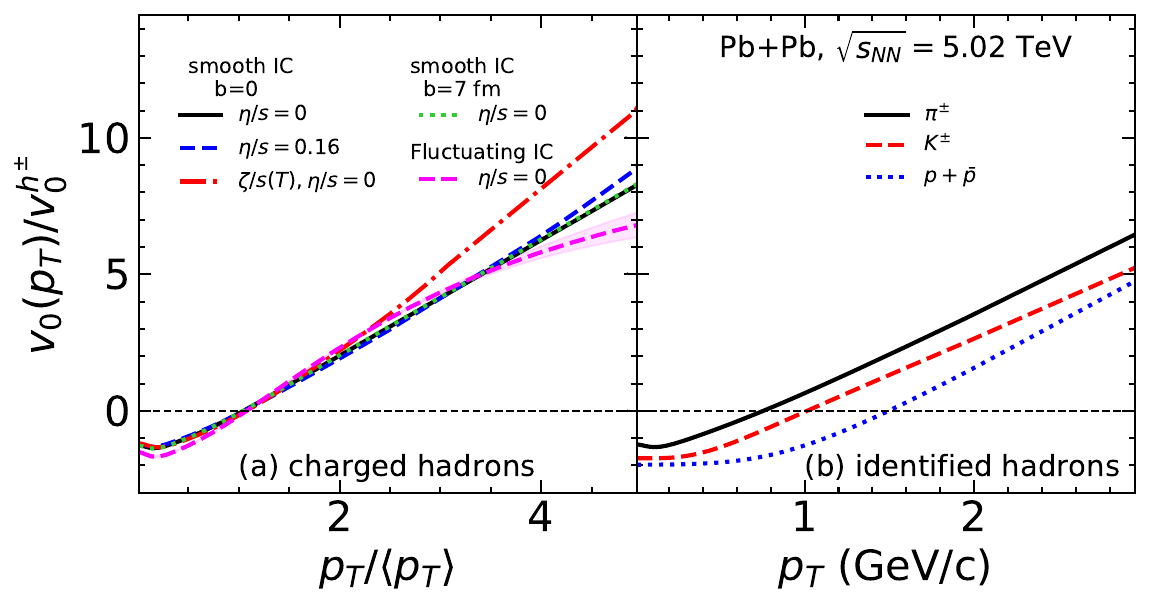}
\caption{Hydrodynamic model predictions of $v_0(p_T)/v_0$ for Pb+Pb collisions at $\sNN = 5.02$ TeV are shown for charged particles in panel (a) and for identified particles in panel (b). In panel (a), the $x$-axis is scaled by $\langle p_T \rangle$. The smooth Initial Condition (IC) results are compared with those from fluctuating IC, as well as between two different impact parameters and for different transport coefficients. In panel (b), results for identified hadrons are shown using smooth IC at $b=0$ fm with $\eta/s=0.16$, with the focus on illustrating the mass hierarchy in $v_0(p_T)$.}
\label{fig1}      
\end{figure}

In Fig.~\ref{fig1} (a), we present the $v_0(p_T)/v_0$ of charged hadrons from both smooth and fluctuating hydrodynamic simulations, which are found to be in good agreement. This indicates that $v_0(p_T)$ predominantly arises from changes in the initial temperature. However, residual effects of fluctuations lead to a small mismatch between the smooth and fluctuating IC. The quantity $v_0(p_T)/v_0$ exhibits a characteristic sign change: it is negative at lower $p_T$ and becomes positive at higher $p_T$. This behavior reflects an anti-correlation between the $p_T$ spectrum and $[p_T]$ in the low-$p_T$ region, while a positive correlation emerges at higher $p_T$.

We present the scaled observable $v_0(p_T)/v_0$, as it is largely independent of system size and centrality at a given collision energy, similar to the scaling observed in $v_n(p_T)/v_n$ \cite{ATLAS:2018ezv}. Furthermore, $\langle p_T \rangle$ itself depends on centrality and transport coefficients, with bulk viscosity notably reducing $\langle p_T \rangle$ for a given centrality class. To eliminate this effect, we rescale the $x$-axis by $\langle p_T \rangle$, which largely removes the dependence on transport coefficients and centrality. The results obtained at $b=7$ fm are consistent with those at $b=0$, further supporting the centrality independence of the scaled observable. While the effect of shear viscosity is minimal, a residual influence of bulk viscosity remains visible in the higher $p_T$ region.

In Fig.~\ref{fig1}(b), we show the results for $v_0(p_T)/v_0$ of identified particles. In this case, the $x$-axis is not scaled by $\langle p_T \rangle$, as our aim is to highlight the mass ordering. A clear mass ordering is observed in the low-$p_T$ region, resembling the behavior seen in other flow coefficients, which is widely considered as a signature of collectivity \cite{ALICE:2022zks}. This mass ordering has also been reported in recent ALICE measurements \cite{ALICE:2025iud}.

\section{$p_T$-cut dependence of $\sigma_{p_T}$: comprehending through $v_0(p_T)$}

\begin{figure}[h]
\centering
\begin{minipage}{0.4\textwidth}
    \centering
    \includegraphics[width=\linewidth]{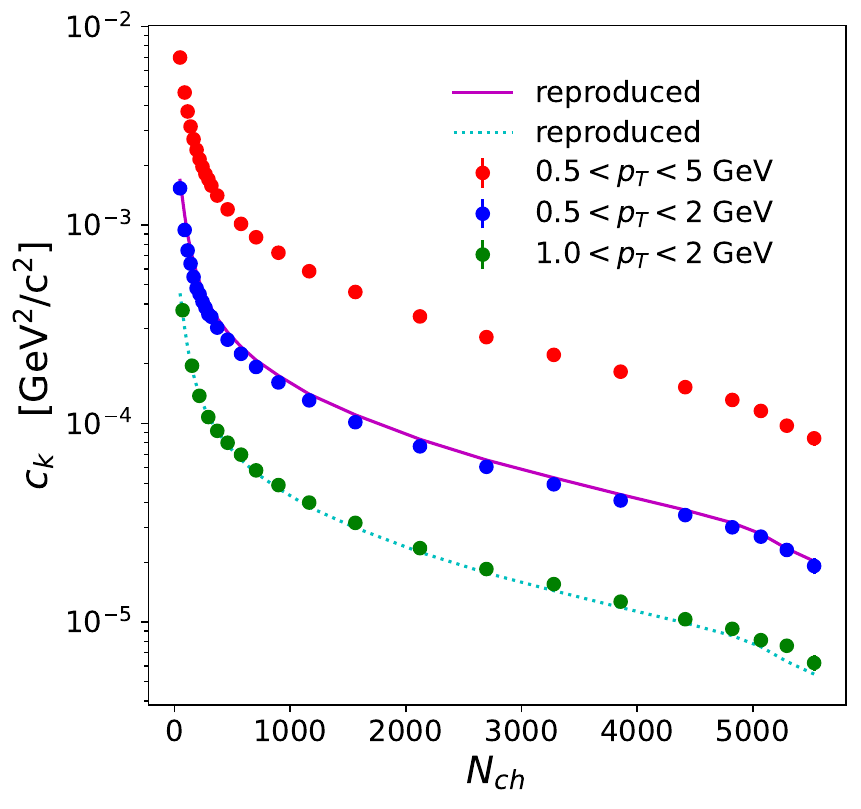}
    \caption{ Symbols show the $N_{\text{ch}}$ dependence of $c_k = \sigma_{p_T}^2$ measured by the ATLAS Collaboration for different $p_T$ cuts in Pb+Pb collisions at $\sNN = 5.02$ TeV \cite{ATLAS:2019pvn}. Using the measured $c_k$ in one $p_T$ window as input (red), predictions for other $p_T$ windows are obtained with hydrodynamic model calculations of $v_0(p_T)$. The lines represent these model predictions.}
    \label{fig:ATLAS_1}
\end{minipage}
\hfill
\begin{minipage}{0.4\textwidth}
    \centering
    \includegraphics[width=\linewidth]{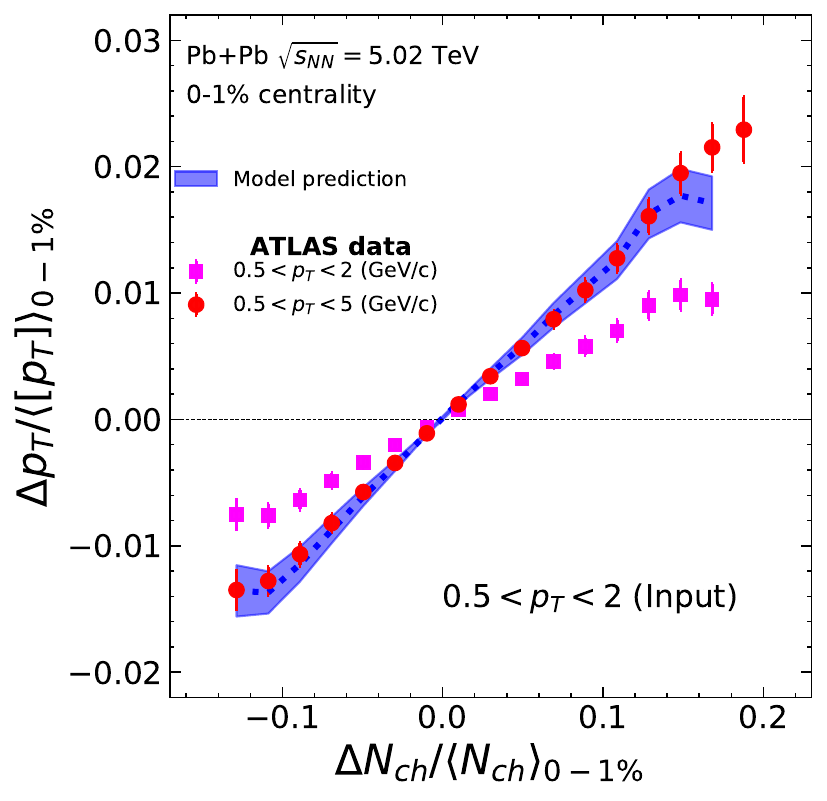}
    \caption{The ratio $\Delta p_T / \langle p_T \rangle$ is shown as a function of $\Delta N_{ch}/\langle N_{ch} \rangle$ for 0–1\% central Pb+Pb collisions at $\sNN = 5.02$ TeV. Experimental data from the ATLAS Collaboration \cite{ATLAS:2024jvf} are compared with our model predictions. We take the window $0.5<p_T<2$ (GeV/c) as input while predicted for the $0.5<p_T<5$ (GeV/c).}
    \label{fig:ATLAS_2}
\end{minipage}
\end{figure}

The dependence of $\sigma_{p_T}$ on the $p_T$ cut, as measured by ATLAS \cite{ATLAS:2019pvn,ATLAS:2024jvf}, is shown in Fig. \ref{fig:ATLAS_1} and \ref{fig:ATLAS_2}. In Fig.~\ref{fig:ATLAS_1}, the quantity $c_k = \langle \left( \langle p_T \rangle -  [p_T] \right)^2 \rangle = \sigma_{p_T}^2$ is plotted as a function of multiplicity ($N_{ch}$) for three different $p_T$ cuts. In Fig.~\ref{fig:ATLAS_2}, the ratio $\Delta p_T / \langle p_T \rangle$, where $\Delta p_T = \langle p_T \rangle - [p_T]$, is presented for 0–1\% central collisions with two different $p_T$ cuts. A strong dependence on the $p_T$ cut is clearly observed in the experimental data, and this behavior can be naturally understood in terms of $v_0(p_T)$.

Since $v_0(p_T)$ is the $p_T$-differential measure of $\sigma_{p_T}$, it describes how the fluctuation is distributed across $p_T$. By integrating $v_0(p_T)$ over the relevant $p_T$ range, one can reproduce the $p_T$ cut dependence of $\sigma_{p_T}$ observed in the ATLAS data. Following a data-driven approach, we first take the measured $\sigma_{p_T}$ in one $p_T$ window and then predict its value for other $p_T$ windows using the $v_0(p_T)$ obtained from hydrodynamic model calculation. The curves shown in both figures represent our model predictions, which exhibit very good agreement with the experimental results.

\section{Summary and Outlook}

The observable $v_0(p_T)$ is a measure of the correlation between the $p_T$ spectra and the event-wise mean transverse momentum, $[p_T]$. It serves as a new probe of the collective behavior of the quark–gluon fluid and offers a direct way to quantify the radial flow of the system. It originates from event-by-event temperature fluctuations of the fluid. We propose the use of the scaled quantity $v_0(p_T)/v_0$ as a function of $p_T/\langle p_T \rangle$, which is found largely independent of centrality and transport coefficients. The $v_0(p_T)$ of identified hadrons exhibits the characteristic mass ordering, further strengthening its interpretation as a manifestation of collectivity. Looking ahead, it would be particularly interesting to explore $v_0(p_T)$ of identified hadrons at higher $p_T$ to investigate whether a meson–baryon splitting occurs, similar to that observed in elliptic and triangular flow coefficients. Another promising direction is to measure $v_0(p_T)$ in small systems, where it could provide valuable insight into the possible existence of collective dynamics. As an application, we have demonstrated that $v_0(p_T)$ can successfully describe the observed $p_T$-cut dependence of $[p_T]$ fluctuations in experimental data.

\section{Acknowledgement}
RS acknowledges the support from the Polish National Science Centre grant 2023/51/B/ST2/01625.

%
%
%

\end{document}